\let\originalcline\cline
\documentclass[pdflatex,sn-mathphys-num,iicol]{sn-jnl}
\let\cline\originalcline

\usepackage{graphicx}%
\usepackage{multirow}%
\usepackage{amsmath,amssymb,amsfonts}%
\usepackage{amsthm}%
\usepackage{mathrsfs}%
\usepackage[title]{appendix}%
\usepackage{xcolor}%
\usepackage{textcomp}%
\usepackage{manyfoot}%
\usepackage{booktabs}%
\usepackage{algorithm}%
\usepackage{algorithmicx}%
\usepackage{algpseudocode}%
\usepackage{listings}%

\usepackage{physics}
\usepackage{subcaption} 

\theoremstyle{thmstyleone}%
\theoremstyle{thmstyletwo}%

\theoremstyle{thmstylethree}%

\begin{document}

\title[Curie--Weiss cell fluid model: Fitting to experimental critical-point data]{Curie--Weiss cell fluid model: Fitting to experimental critical-point data}


\author*{\fnm{I.~V.} \sur{Pylyuk}}\email{piv@icmp.lviv.ua}

\author{\fnm{O.~A.} \sur{Dobush}}\email{dobush@icmp.lviv.ua}

\author{\fnm{M.~P.} \sur{Kozlovskii}}\email{mpk@icmp.lviv.ua}

\author{\fnm{R.~V.} \sur{Romanik}}\email{romanik@icmp.lviv.ua}

\author{\fnm{M.~A.} \sur{Shpot}}\email{shpot.mykola@gmail.com}
\affil{\orgname{Yukhnovskii Institute for Condensed Matter Physics of the National Academy of Sciences of Ukraine}, \orgaddress{
		\city{Lviv}, \postcode{79011}, \country{Ukraine}}}


\abstract{We examine how the Curie--Weiss cell fluid model can be quantitatively related to real fluids by determining its microscopic parameters from experimental critical-point data for water, carbon dioxide, and three polyol ester lubricants (POE5, POE7, and POE9). The microscopic parameters of the model are determined from the experimental values of the critical temperature and critical density: the global attraction coupling constant is fixed by the critical temperature, and the cell volume by the critical density. The critical pressure then follows as an independent model prediction, deviating from experimental values by 10--41$\%$ across the studied substances. Varying the ratio of local repulsion to global attraction $f=J_2/J_1$ over a wide range has only a weak effect on this deviation. For $f\geq1$, the model's critical compressibility factor lays within a narrow range $Z_c\approx0.382-0.386$, remaining close to the mean-field van der Waals value $Z_c=3/8=0.375$ and consistently overestimating experimental values. The best agreement with experiment is obtained for the polyol ester lubricants and improves along the POE5--POE7--POE9 series with increasing molar mass, suggesting that multiple-occupancy cell fluid models are well suited to systems of large molecules, particularly soft-matter systems with bounded interaction potentials. We further show that generalizing the model by introducing a temperature- and phase-dependent effective attraction allows it to describe a triple point and to reproduce the broad range of critical-to-triple temperature ratios $T_c/T_{\mathrm{tr}}$ characteristic of real fluids.}

\keywords{cell fluid model, Curie--Weiss interaction, experimental data, critical point, triple point}



\maketitle

\section{Introduction}

One of the central problems in the statistical physics of fluids is establishing the relationship between the microscopic characteristics of interparticle interactions and the macroscopic properties of matter. Of particular interest in this context are critical and triple points, which are fundamental reference points of the phase diagram. Experimental values of temperature, mass density, and pressure at the critical point can be used to determine the microscopic parameters of a theoretical model. At the same time, the characteristics of the triple point provide additional information about the phase behavior of the system at low temperatures and can be used to analyze the mechanisms leading to three-phase equilibrium.

A wide range of theoretical approaches has been developed to describe phase transitions in fluids, from the classical van der Waals theory to modern renormalization-group methods and computer simulations~\cite{Johnston14,Goldenfeld18,Fisher98,FrenkelSmit23,AllenTildesley17,BinderHeermann19}. The relationship between interparticle interactions, fluid structure, and phase behavior is analyzed in detail in the classical books~\cite{Stanley71,HansenMcDonald13}. Mean-field theories remain an important tool for studying phase transitions, since they allow one to establish the main features of phase behavior and, in many cases, to obtain analytical results. At the same time, it is well known that such theories do not take into account spatial density correlations and critical fluctuations, which determine the universal properties of real systems near the critical point~\cite{Stanley71,Fisher98}.

Despite significant progress in the theory of liquids, the problem of constructing and studying models that combine analytical solvability, a clear physical interpretation of the parameters, and the possibility of direct comparison with experimental data remains relevant. The cell fluid model with Curie--Weiss interaction considered in this work belongs to this class of models. This model has an analytical solution~\cite{KKD18,KKD20,KD22} and exhibits a sequence of first-order phase transitions. The influence of the microscopic parameters -- the cell volume $v$ and the ratio of the interaction constants $f$ -- on the critical behavior of the Curie--Weiss cell model was studied in Refs.~\cite{PDKRS25,DKPP26}. It was shown that the reduced critical-point coordinates do not depend on $v$, except for the chemical potential. At the same time, these coordinates depend on the parameter $f$, which can be regarded as a tuning parameter for them. In Refs.~\cite{DS24arxiv,DKPRS26}, analytical results were derived for some special cases of the model. Moreover, it was recently shown that introducing effective temperature-dependent attractive interactions qualitatively changes the phase behavior of the model, leading to the appearance of a triple point in the phase diagram~\cite{KDPRS26}. We also used the cell model in previous studies~\cite{KD16,KDP17,KPD18,Pylyuk20,PKR26}, where the interparticle interactions were described by the Morse potential.

An important aspect of applying the model to the description of real substances is the quantitative determination of its microscopic parameters from experimental data. This approach makes it possible to establish a relation between the model parameters and the characteristics of specific physical systems.

Whereas determining the microscopic parameters from critical-point data connects the model directly to macroscopic observables, analyzing the triple point allows us to evaluate the impact of temperature-dependent attraction on the phase diagram.

Carbon dioxide, water, and polyol ester lubricants were selected for the study. These substances differ significantly in molar mass, molecular structure, the character of intermolecular interactions, and critical parameters. Carbon dioxide is a simple molecular fluid with relatively weak intermolecular interactions, water is a strongly associated liquid with hydrogen bonds, whereas polyol ester lubricants are organic liquids with substantially larger molar masses and more complex molecular structures. This choice makes it possible to assess the applicability of the proposed approach to determining the microscopic model parameters for a broad class of liquids. Water is particularly illustrative in this context because it exhibits complex thermodynamic behavior and is characterized by a number of anomalous properties and special temperatures~\cite{BRC23,BRC25}. Reliable reference data on the critical-point coordinates required to determine the model parameters are available for all selected substances.

We focus on the model's lowest-temperature critical point, interpreting it as a liquid--gas critical point. Since the present mean-field Curie--Weiss model does not capture local ordering, our identification of phases rests purely on a density-based classification.

The aim of this work is to determine the microscopic parameters of the Curie--Weiss cell model from experimental critical-point data for real fluids and to assess the agreement between the model and experiment. Particular attention is paid to the critical compressibility factor $Z_c$ as a dimensionless measure of this agreement. Since the basic model does not describe a triple point, we also consider its generalization with a phase-dependent, temperature-dependent attractive interaction, while keeping the parameters of the basic model determined from the critical point unchanged. For the generalized model, we investigate how the parameters governing the temperature dependence of the effective attraction affect the position of the triple point and the ratio $T_c/T_{\rm tr}$.

\section{Cell model of a fluid}\label{sec2}
As a model of a fluid, we consider an open system of point particles confined to a bounded region of three-dimensional space ${\mathcal V} \subset \mathbb R^3$ with volume $V = \abs{\mathcal V}$. The region $\mathcal V$ is divided into $N_v$ identical non-overlapping cubic cells $\Delta_l$, $l\in\{1,...,N_v\}$. Each cell has volume $v=c^3$, where $c$ is its linear size. Thus, the total volume of the system is $V=v N_v$, and in the thermodynamic limit, $V \to \infty$, $N_v \to \infty$, the cell volume remains constant.

The pair interaction potential $\Psi_{N_v} (\vb{r}, \vb{r}')$ between particles with coordinates
$\vb{r}$ and $\vb{r}'$ is given by
\begin{equation}
	\Psi_{N_v} (\vb{r}, \vb{r}')=- \frac{J_1}{N_v} + J_2 \sum_{l =1}^{N_v}
	I_{\Delta_l}(\vb{r}) I_{\Delta_l}(\vb{r}').
	\label{2d1et}
\end{equation}
The first term in~\eqref{2d1et} describes the global attraction between all particle pairs in the system and is characterized by the coupling constant $J_1>0$. The factor $1/N_v$ corresponds to the standard Curie--Weiss scaling and ensures the extensivity of the interaction energy in the thermodynamic limit. The second term, characterized by the constant $J_2>0$, describes the local repulsion between particles located in the same cell $\Delta_l$. The interaction defined by Eq.~\eqref{2d1et} will be referred to as the Curie--Weiss interaction~\cite{KKD20,KD22}.
The indicator function $I_{\Delta_l}(\vb{r})$ is defined as
\begin{equation}
	I_{\Delta_l}(\vb{r})=\left\{
	\begin{array}{l}
		1, \vb{r}\in \Delta_l , \\
		0, \vb{r}\notin \Delta_l.
	\end{array}
	\right.
	\label{2d2et}
\end{equation}
The interaction potential~\eqref{2d1et} satisfies the Ruelle thermodynamic stability condition~\cite{Ruelle70} if $J_2 \geq J_1>0$, with $J_2/J_1=1$ corresponding to the boundary of the thermodynamically stable region~\cite{DKPRS26}. The particle positions in space remain continuous, while the division of the system into cells is used only to define the local part of the interaction. The number of particles in each cell is not restricted; that is, states with multiple cell occupancy are allowed.

The primary microscopic parameters governing the interaction and phase behavior of the model are the cell volume $v$ and the coupling constant $J_1$. Using $v$ and $J_1$ as natural scales of volume and energy, we describe the system in terms of dimensionless (reduced) thermodynamic variables
\begin{equation}
	\label{def_reduced}
	T^* = k_{\rm B} T / J_1, \quad P^* = P v / J_1, \quad n^* = n v.
\end{equation}
Here, $T^*$ is the reduced temperature, $P^*$ is the reduced pressure, and $n^*$ is the reduced number density of particles. We use the standard notation for temperature, $T$, pressure, $P$, particle number density, $n = N/V$, and the Boltzmann constant, $k_{\rm B} = 1.380649 \times 10^{-23} \, \text{J} \cdot \text{K}^{-1}$. The quantity $N$ denotes the total number of particles in the system. When calculations are performed for the model within the grand canonical ensemble formalism, $n = \langle N \rangle / V$, where $\langle N \rangle$ is the mean number of particles in the system. The phase behavior of the model in terms of the reduced thermodynamic quantities is determined by the ratio $f=J_2/J_1$, and does not depend on the value of the cell volume $v$~\cite{RDKPS26a}.

The model under consideration has an infinite sequence of critical points, each specified by a point in temperature--number density--pressure coordinates~\cite{KKD20,KD22}. In this work, we focus on the critical point with the lowest density. In applying the model to real fluids, the critical point with the lowest density is identified with the liquid--gas (liquid--vapor) critical point. Accordingly, the lower-density phase is interpreted as the gas phase, while the higher-density phase is interpreted as the liquid phase. Thus, the critical point of the model is specified by three coordinates $(T^*_{\rm c}; n^*_{\rm c}; P^*_{\rm c})$. The three quantities commonly used to specify the coordinates of the critical point of real substances are temperature $T$, pressure $P$, and mass density $\rho = m/V$, where $m$ is the mass of the entire system. The relation between the number density $n$ and the mass density $\rho$ is given by
\begin{equation}
	\label{n_vs_rho}
	n = \frac{\rho N_{\rm A}}{M_{\rm mol}},	
\end{equation}
where $N_{\rm A} = 6.02214076 \times 10^{23} \, \text{mol}^{-1}$ is Avogadro's number, and $M_{\rm mol}$ is the molar mass of the substance. The critical values of the dimensional quantities, $(T_c; \rho_c; P_c)$, are different for each substance and are determined experimentally.

\section{Comparison with experimental data for specific substances}
\label{sec3}
The relation between the cell model and a particular substance can be established by matching the critical-point coordinates. For a fixed value of $f=J_2/J_1$, the dimensionless critical-point coordinates of the model are known. Then, knowing two experimental critical-point coordinates for a given substance, we can determine the two dimensional model parameters $J_1$ and $v$.

For example, if the experimental values of the critical temperature $T_c$ and mass density $\rho_c$ are known, along with the corresponding model quantities $T^*_c$ and $n^*_c$, then the parameters $J_1$ and $v$ follow from Eqs.~\eqref{def_reduced} and~\eqref{n_vs_rho}:
\begin{equation}
	\label{cond-T-n}
	J_1 = \frac{k_{\rm B} T_c}{T^*_c}, \quad v = v_D \equiv \frac{M_{\rm mol} n^*_c}{N_{\rm A} \rho_c}.
\end{equation}
Given $J_1$ and $v$, the theoretical value of the critical pressure is calculated as
\begin{equation}
	\label{theor_P}
	P^{\rm (model)}_{\rm c} = \frac{P^*_{\rm c} J_1}{v_D},
\end{equation}
and can be compared with the corresponding experimental value.

Similarly, another pair of quantities can be used -- the critical temperature and pressure. From the experimental values $T_c$ and $P_c$ we obtain the corresponding model quantities $T^*_c$ and $P^*_c$
\begin{equation}
	\label{cond-T-P}
	J_1 = \frac{k_{\rm B} T_c}{T^*_c}, \quad v = v_P \equiv \frac{P^*_c}{P_c}\frac{k_{\rm B} T_c}{T^*_c},
\end{equation}
and the theoretical value of the critical density
\begin{equation}
	\label{theor_rho}
	\rho^{\rm (model)}_c = \frac{M_{\rm mol} n^*_c}{N_{\rm A} v_P}.
\end{equation}
This value, in turn, can be compared with the experimental critical density.

Next, we consider both variants. To distinguish between the two ways of determining the cell volume, we use the notations $v_D$ and $v_P$. Condition~\eqref{cond-T-n} will be referred to below as the density-matching condition, and condition~\eqref{cond-T-P} as the pressure-matching condition. In principle, another pair of experimental thermodynamic quantities can also be used to determine the two dimensional model parameters $J_1$ and~$v$.

The comparison with experimental data will be performed for two one-component fluids -- carbon dioxide (CO$_2$) and water (H$_2$O)
-- as well as for the polyol ester lubricants POE5 (C$_{25}$H$_{44}$O$_8$), POE7 (C$_{33}$H$_{60}$O$_8$), and POE9 (C$_{41}$H$_{76}$O$_8$). The main difference between the POE5, POE7, and POE9 lubricants lies in the chemical composition of the base ester and in their kinematic viscosity. The higher the lubricant number, the greater its kinematic viscosity at operating temperatures. The chemical difference between POE5, POE7, and POE9 is related to the molecular structure of the alcohols and acids from which these esters are synthesized. Though all of them are polyol esters (formed by the reaction of polyhydric alcohols with fatty acids), the difference in their number is determined by the length and branching of the chains. The higher the number, the more complex and heavier the ester molecule.

We use the available experimental data for the coordinates of the liquid--vapor critical points of CO$_2$~\cite{SW96,SW03c} and
H$_2$O~\cite{GreenSouthard19,LBHM25,miniREFPROP20,BWQL14}, as well as the reference critical parameters for POE5, POE7, and POE9~\cite{BFHetAl19}.
The critical and molecular characteristics of these substances are given in Table~\ref{tab_1et}.

\begin{table*}[htbp]
	\centering
	\noindent\caption{Critical parameters, molar masses, and molecular masses of the substances studied}\vskip3mm\tabcolsep4.5pt
	\label{tab_1et}
	
	\noindent{\footnotesize\begin{tabular}{|l|c|c|c|c|c|}
			\hline \multicolumn{1}{|c}
			{\rule{0pt}{5mm}Quantity} & \multicolumn{1}{|c}{CO$_2$} &
			\multicolumn{1}{|c}{H$_2$O} &
			\multicolumn{1}{|c}{POE5} &
			\multicolumn{1}{|c}{POE7} &
			\multicolumn{1}{|c|}{POE9}\\[2mm]%
			\hline%
			\rule{0pt}{5mm}Critical temperature $T_c \, (\text{K})$ & 304.1 & 647.1 &
			890.0 & 940.0 & 970.0\\%
			Critical mass density $\rho_c \, (\text{kg/m}^3)$ & 467.6 & 322.0 &
			262.8 & 241.0 & 220.3\\%
			Critical pressure $P_c \, (\times 10^6\,\text{Pa})$ & 7.38 & 22.06 &
			1.270 & 1.030 & 0.885\\%
			Molar mass $M_{mol} \, (\text{kg/mol})$ & 0.044 & 0.018 &
			0.473 & 0.585 & 0.697\\%
			Molecular mass $m_0 \, (\times 10^{-26}\,\text{kg})$ & 7.3 & 3.0 &
			78.5 & 97.1 & 115.8\\[2mm]%
			\hline
	\end{tabular}}
\end{table*}

Thus, we proceed to determine the microscopic parameters of the cell fluid model with Curie--Weiss interaction using the conditions of agreement with experimental data. Using the density-matching condition~\eqref{cond-T-n}, we calculate the theoretical values of the critical pressure, while using the pressure-matching condition~\eqref{cond-T-P}, we calculate the theoretical values of the critical density. We also analyze the agreement between model predictions and experimental data at critical points using the compressibility factor $Z_c$, examining the influence of the parameter $f$ on the model's ability to match experimental values. Finally, we extend our analysis to the triple point and draw our main conclusions.

\begin{table}[htbp]
	\centering
	\noindent\caption{Microscopic parameters of the cell model calculated for different substances. Here, $v_D$ is the cell volume calculated from~\eqref{cond-T-n}, and $v_P$ is the cell volume calculated from~\eqref{cond-T-P}. The numerical values are obtained for $f = 1.5$}\vskip3mm\tabcolsep4.5pt
	\label{tab2_micro_params}
	
	\noindent{\footnotesize\begin{tabular}{|c|c|c|c|}
			\hline
			{\rule{0pt}{5mm}Substance} & $J_1 \, (\times 10^{-20} \, \text{J})$ & $v_P \, (\text{\AA}^3)$ &
			$v_D \, (\text{\AA}^3)$\\[2mm]%
			\hline%
			\rule{0pt}{5mm}CO$_2$ & 1.67 & 110.6 & 78.7 \\%
			H$_2$O & 3.56 & 78.7 & 46.8 \\%
			POE5 & 4.89 & 1881.2 & 1505.9 \\%
			POE7 & 5.16 & 2449.9 & 2031.0 \\%
			POE9 & 5.33 & 2942.3 & 2647.2 \\[2mm]%
			\hline
	\end{tabular}}
\end{table}

Let us fix the parameter $f$ at $f=1.5$. In this case, the critical-point coordinates of the model are~\cite{KDPRS26}:
\begin{equation}
	\label{cr_point}
	T^*_c = 0.2513, \quad n^*_c = 0.5039, \quad P^*_c = 0.04886.
\end{equation}
The values of the microscopic model parameters $J_1$ and $v$ for the five substances listed in Table~\ref{tab_1et} are given in Table~\ref{tab2_micro_params}.
As can be seen from Table~\ref{tab2_micro_params}, the values of $v$ obtained from the pressure-matching condition are always larger than the corresponding values obtained from the density-matching condition, $v_P > v_D$.

\begin{table}[htbp]
	\centering
	\noindent\caption{Critical-point pressure. Here, $P_c$ is the experimental value, and $P^{\rm (model)}_{c}$ is the pressure calculated from the density-matching condition~\eqref{cond-T-n}. The calculations are performed for $f = 1.5$}\vskip3mm\tabcolsep4.5pt
	\label{tab_theor_P}
	
	\noindent{\footnotesize\begin{tabular}{|c|c|c|c|}
			\hline \multicolumn{1}{|c}
			{\rule{0pt}{5mm}Substance} & \multicolumn{1}{|c}{$P_c \, (\times 10^6\,\text{Pa})$} &
			\multicolumn{1}{|c}{$P^{\rm (model)}_{c} \, (\times 10^6\,\text{Pa})$} &
			\multicolumn{1}{|c|}{$\delta_P$}\\[2mm]%
			\hline%
			\rule{0pt}{5mm}CO$_2$ & 7.38 & 10.37 & $\sim$29\%\\%
			H$_2$O & 22.06 & 37.14 & $\sim$41\%\\%
			POE5 & 1.270 & 1.587 & $\sim$20\%\\%
			POE7 & 1.030 & 1.242 & $\sim$17\%\\%
			POE9 & 0.885 & 0.984 & $\sim$10\%\\[2mm]%
			\hline
	\end{tabular}}
\end{table}

\begin{table}[htbp]
	\centering
	\noindent\caption{Mass density at the critical point. Here, $\rho_c$ is the experimental value, and $\rho^{\rm (model)}_{c}$ is the density calculated from the pressure-matching condition~\eqref{cond-T-P}. The calculations are performed for $f = 1.5$}\vskip3mm\tabcolsep4.5pt
	\label{tab_theor_rho}
	
	\noindent{\footnotesize\begin{tabular}{|c|c|c|c|}
			\hline \multicolumn{1}{|c}
			{\rule{0pt}{5mm}Substance} & \multicolumn{1}{|c}{$\rho_c \, (\text{kg} \, \text{m}^{-3})$} &
			\multicolumn{1}{|c}{$\rho^{\rm (model)}_{c} \, (\text{kg} \, \text{m}^{-3})$} &
			\multicolumn{1}{|c|}{$\delta_D$}\\[2mm]%
			\hline%
			\rule{0pt}{5mm}CO$_2$ & 467.6 & 332.8 & $\sim40\%$\\
			H$_2$O & 322.0 & 191.3 & $\sim68\%$\\%
			POE5 & 262.8 & 210.4 & $\sim25\%$\\%
			POE7 & 241.0 & 199.8 & $\sim21\%$ \\
			POE9 & 220.3 & 198.2 & $\sim11\%$\\[2mm]%
			\hline
	\end{tabular}}
\end{table}

Now, using the obtained values of the microscopic parameters, we calculate the theoretical pressure~\eqref{theor_P} using the density-matching condition, and the theoretical critical density~\eqref{theor_rho} using the pressure-matching condition. The results are given in Tables~\ref{tab_theor_P} and~\ref{tab_theor_rho}, respectively. To quantify the discrepancies between the experimental and theoretical results, we use the relative deviation:
\begin{equation}
	\delta_P = \frac{\abs{P_c - P^{(model)}_c}}{P^{(model)}_c} \cdot 100\%,
\end{equation}
for the pressure in Table~\ref{tab_theor_P}, and
\begin{equation}
	\delta_D = \frac{\rho_c - \rho^{(model)}_c}{\rho^{(model)}_c} \cdot 100\%
\end{equation}
for the density in Table~\ref{tab_theor_rho}.

\emph{Choice of the cell volume.} Combining equations~\eqref{cond-T-n}--\eqref{theor_rho} yields the following relation between the cell volumes $v_P$ and $v_D$ used in the two schemes above:
\begin{equation}
	\frac{v_P}{v_D} = \frac{P_c^{(model)}}{P_c} = \frac{\rho_c}{\rho_c^{(model)}}.
\end{equation}
n subsequent calculations, we use the cell volume determined from the critical-density matching condition (5),
\begin{equation}
	v = v_D.
\end{equation}
With this choice, the critical temperature and density are reproduced exactly by construction, whereas the critical pressure is an independent prediction of the model.

\subsection{Dependence of the critical pressure agreement on the parameter~$f$}
\label{sec5}

Table~\ref{tab_3et} compares the experimental critical pressures of the considered substances with theoretical values from Eq.~\eqref{theor_P} for several choices of the parameter $f$. Specifically, we consider $f=1.0$, representing the thermodynamic stability boundary of the model~\cite{DS24arxiv,DKPRS26}, an intermediate value $f=2.0$, and $f=10.0$, described by the analytical solutions of~\cite[Sec.~9]{DKPRS26} in the large-$f$ limit. Analogous results for $f=1.5$ are listed in Table~\ref{tab_theor_P}.

\begin{table*}[htbp]
	\centering
	\noindent\caption{Comparison of the calculated and experimental critical pressures for the selected substances at different values of the parameter $f$}\vskip3mm\tabcolsep4.5pt
	\label{tab_3et}
	
	\noindent{\footnotesize\begin{tabular}{|c|c|c|c|c|c|c|c|}
			\hline%
			\multicolumn{1}{|c}{\rule{0pt}{5mm}Substance} &
			\multicolumn{1}{|c}{$P_c \, (\times 10^6\,\text{Pa})$} & \multicolumn{2}{|c}{$f = 1.0$} &
			\multicolumn{2}{|c}{$f = 2.0$}  & \multicolumn{2}{|c|}{$f = 10.0$}\\[2mm]%
			\cline{3-8}%
			&& \multicolumn{1}{c}{\rule{0pt}{5mm}$P^{(model)}_{c} \, (\times 10^6\,\text{Pa})$} &
			\multicolumn{1}{|c}{$\delta_P$} &
			\multicolumn{1}{|c}{$P^{(model)}_{c} \, (\times 10^6\,\text{Pa})$} &
			\multicolumn{1}{|c}{$\delta_P$} &
			\multicolumn{1}{|c}{$P^{(model)}_{c} \, (\times 10^6\,\text{Pa})$} &
			\multicolumn{1}{|c|}{$\delta_P$}\\[2mm]%
			\hline%
			\rule{0pt}{5mm}CO$_2$ & 7.38 &
			10.27 & $\sim$28\% & 10.38 & $\sim$29\% &
			10.38 & $\sim$29\%\\%
			H$_2$O & 22.06 &
			36.80 & $\sim$40\% & 37.17 & $\sim$41\% &
			37.18 & $\sim$41\%\\%
			POE5 & 1.270 &
			1.572 & $\sim$19\% & 1.588 & $\sim$20\% &
			1.588 & $\sim$20\%\\%
			POE7 & 1.030 &
			1.231 & $\sim$16\% & 1.244 & $\sim$17\% &
			1.244 & $\sim$17\%\\%
			POE9 & 0.885 &
			0.975 & $\sim$9\% & 0.985 & $\sim$10\% &
			0.985 & $\sim$10\%\\[2mm]%
			\hline
	\end{tabular}}
\end{table*}

\emph{Choice of the parameter $f$.} As shown in Table~\ref{tab_3et}, the deviation between predicted and experimental critical pressures for all substances changes only by approximately 1\% as $f$ varies from 1 to 10. At $f = 1.5$, this deviation is nearly the same as for larger values of $f$ (see Tables~\ref{tab_3et} and~\ref{tab_theor_P}). In subsequent calculations, we set $f=1.5$, the same as in Ref.~\cite{KDPRS26}, where the triple point was described in an extended model with an effective temperature-dependent attraction.

As seen from Tables~\ref{tab_theor_P} and~\ref{tab_3et}, the deviation between the experimental and model critical pressures decreases systematically along the sequence H$_2$O -- CO$_2$ -- POE5 -- POE7 -- POE9, i.e., with increasing molar mass. The best agreement is achieved for POE9, whereas the largest deviation is found for H$_2$O. This tendency suggests that the mean-field description becomes more suitable for systems composed of larger molecules. The pronounced deviation for water can be attributed to the specific character of its intermolecular interactions, in particular hydrogen bonding and spatial correlations, which are outside the scope of the present mean-field model.

\subsection{Some auxiliary microscopic quantities of the model}

\begin{table*}[htbp]
	\centering
	\noindent\caption{Comparative data for the calculated microscopic quantities of the model for five substances (CO$_2$, H$_2$O, POE5,
		POE7, POE9)}\vskip3mm\tabcolsep4.5pt
	\label{tab_5et}
	
	\noindent{\footnotesize\begin{tabular}{|c|c|c|c|c|c|}
			\hline \multicolumn{1}{|c}
			{\rule{0pt}{5mm}Quantity} & \multicolumn{1}{|c}{CO$_2$} &
			\multicolumn{1}{|c}{H$_2$O} &
			\multicolumn{1}{|c}{POE5} &
			\multicolumn{1}{|c}{POE7} &
			\multicolumn{1}{|c|}{POE9}\\[2mm]%
			\hline%
			\rule{0pt}{5mm}$J_1 \, (\times 10^{-20}\,\text{J})$ & 1.67 & 3.56 &
			4.89 & 5.16 & 5.33\\%
			$J_2 \, (\times 10^{-20}\,\text{J})$ & 2.51 & 5.33 &
			7.33 & 7.75 & 7.99\\%
			$v \, (\text{\AA}^3)$ & 78.7 & 46.8 &
			1505.9 & 2031.0 & 2647.2\\%
			$c \, (\text{\AA})$ & 4.3 & 3.6 &
			11.5 & 12.7 & 13.8\\%
			$n_c \, (\times 10^{-3}\,\text{\AA}^{-3})$ & 6.4 & 10.8 &
			0.33 & 0.25 & 0.19\\%
			$\Lambda_{J} \, (\text{\AA})$ & 0.076 & 0.081 &
			0.013 & 0.012 & 0.011\\%
			$\Lambda \, (\text{\AA})$ & 0.151 & 0.161 &
			0.027 & 0.024 & 0.021\\%
			$v^* \, (\times 10^9)$ & 0.00018 & 0.000088 &
			0.61 & 1.2 & 2.2\\%
			$n_c v$ & $\approx$0.5 & $\approx$0.5 &
			$\approx$0.5 & $\approx$0.5 & $\approx$0.5\\[2mm]%
			\hline
	\end{tabular}}
\end{table*}

When considering the cell fluid model with Curie--Weiss interaction in the grand canonical ensemble~\cite{RDKPS26a}, several auxiliary quantities arise. These are the coupling constant of the repulsive interaction, $J_2 = f J_1$, the linear size of the cell, $c = v^{1/3}$, the particle number density at the critical point, $n_c = \rho_c N_{\rm A} / M_{\rm mol}$, the thermal de Broglie wavelength at the critical point, $\Lambda_c = (2\pi \hbar^2 / (m_0 k_{\rm B} T_c))^{1/2}$, and $\Lambda_J = (2\pi \hbar^2 / (m_0 J_1))^{1/2}$, where $m_0$ is the mass of one molecule. Another quantity that arises in the calculations~\cite{RDKPS26a} is the ratio of the cell volume to $\Lambda_J^3$, $v^*=v / \Lambda^3_J$. All these quantities are given in Table~\ref{tab_5et} for the five substances studied.

The calculated values of the global attraction coupling constant $J_1$ and the local repulsion coupling constant $J_2$ for the five substances (CO$_2$, H$_2$O, POE5, POE7, POE9) both increase with rising critical temperature. For the polyol ester lubricants, the values of $J_1$ and $J_2$ grow systematically with molar mass.

For small molecules (H$_2$O, CO$_2$), the cell volume $v$ is of the order of 50 -- 80 $\text{\AA}^3$, with the smallest value for water, whereas for the polyol ester lubricants POE5, POE7, and POE9 it reaches thousands of $\text{\AA}^3$. The linear cell size $c = (\rho_c^*/n_c)^{1/3} \approx (0.5/n_c)^{1/3}$ is typical for intermolecular distances: for CO$_2$ and H$_2$O it corresponds to several angstroms, whereas for POE5, POE7, and POE9 it extends to the nanometer scale.

The critical number density $n_c$ is the number of particles (molecules) per unit volume at the critical point. H$_2$O
has the highest critical number density, while CO$_2$ has an intermediate value. The values of $n_c$ for POE5, POE7, and POE9 are two orders of magnitude lower than that for H$_2$O.

The dimensionless volume $v^* = v/\Lambda_{J}^3$
compares the cell volume $v$ against the quantum thermal length scale $\Lambda_{J}^3$.
For $v^* \gg 1$ quantum effects can be neglected and the system is well described by classical statistical mechanics. For CO$_2$ and H$_2$O, we obtain $v^* \sim 10^5$, and for POE5, POE7, and POE9, $v^* \sim 10^9$. This confirms that our classical mean-field description of the selected substances is legitimate, which is much more pronounced for heavy-molecule lubricants.

\section{Evaluating the agreement between model and experiment via the critical compressibility factor}
\label{sec7}

The critical compressibility factor is an important dimensionless characteristic of the critical point and is defined by the relation
\begin{equation}
	Z_c = \frac{P_c}{n_c k_B T_c}.
	\label{7d3et}
\end{equation}
The quantity $Z_c$ is the ratio of the actual critical pressure to the ideal-gas pressure at the same temperature $T_c$ and number density $n_c$. Using relation~\eqref{n_vs_rho} between the number density and mass density, the experimental values for $Z_c$ can be calculated via
\begin{equation}
	\label{Zexp}
	Z^{(exp)}_c = \frac{P_c M_{\rm mol}}{\rho_c N_{\rm A} k_{\rm B} T_c}.
\end{equation}
For the cell model, the corresponding quantity is expressed in terms of the reduced critical-point coordinates:
\begin{equation}
	\label{Ztheor}
	Z^{(model)}_{c} = \frac{P^*_c}{n^*_c T^*_c}.
\end{equation}
Thus, for a fixed value of $f=J_2/J_1$, the quantity $Z_c^{(\mathrm{model})}$ does not depend on the scale parameters $J_1$ and $v$ and is the same for all substances with which the model is compared. For $f=1.5$, using the values in~\eqref{cr_point} we obtain
\begin{equation}
	Z_c^{(\mathrm{model})} \simeq 0.386.
\end{equation}
The critical compressibility factor can be related to the two ways of determining the cell volume considered in Section~\ref{sec3}. From Eqs.~\eqref{cond-T-n}, \eqref{cond-T-P}, \eqref{Zexp} and \eqref{Ztheor}, it follows that
\begin{equation}
	\frac{v_P}{v_D} = \frac{Z_c^{(\mathrm{model})}}{Z_c^{(\mathrm{exp})}}.
\end{equation}

When the density-matching condition is used, $v=v_D$, the critical temperature and density are reproduced exactly. Then, for this choice,
\begin{equation}
	\frac{P_c^{(model)}}{P_c} = \frac{Z_c^{(\mathrm{model})}}{Z_c^{(\mathrm{exp})}},
\end{equation}
the discrepancy in the critical pressure discussed in Section~\ref{sec3} directly determines the discrepancy in the critical compressibility factors. In this sense, the analysis of $Z_c$ is not an independent additional test, but provides a dimensionless formulation of the matching condition for the three critical-point coordinates.

The results of comparing the experimental values of the critical compressibility factor for the five substances considered with the corresponding model value are given in Table~\ref{tab_7et}. To quantify the discrepancy between the experimental and model values, we use the relative deviation
\begin{equation}
	\delta_Z = \frac{\abs{Z^{(exp)}_c - Z^{(model)}_c}}{Z^{(model)}_c} \cdot 100\%.
\end{equation}
For the considered substances, the experimental values of $Z_c^{(\mathrm{exp})}$ range from $0.229$ to $0.347$ (see Table~\ref{tab_7et}), whereas for $f=1.5$ the model gives $Z_c^{(\mathrm{model})}\simeq0.386$. Thus, for all five substances
\begin{equation}
	Z_c^{(\mathrm{model})} > Z_c^{(\mathrm{exp})},
\end{equation}
which reflects a systematic overestimation of the critical pressure when $v=v_D$ is chosen. For many simple molecular fluids, typical values of $Z_c$ lie in the range $0.23$?$0.30$~\cite[Table~5.1]{Stanley71}. For the polyol esters considered here, the values of $Z_c^{(\mathrm{exp})}$ are higher and approach the model value.

The question arises whether this discrepancy can be eliminated by adjusting the parameter $f=J_2/J_1$. The calculated critical-point coordinates for different values of $f$ are given in Table~\ref{tab_8et}. In the thermodynamically stable region $f\ge1$, the model compressibility factor varies only within a narrow interval
\begin{equation}
	0.382373 \leq Z_c^{(\mathrm{model})} (f) \leq 0.386294.
\end{equation}
Thus, even the minimum theoretical value, attained near $f=1$, exceeds $Z_c^{(\mathrm{exp})}$ for all considered substances. Therefore, no physically admissible value of $f\ge1$ can satisfy the condition $Z_c^{(\mathrm{model})} = Z_c^{(\mathrm{exp})}$.

As $f$ increases, the critical temperature $T_c^*$ and the critical number density $n_c^*$ decrease slightly, whereas $Z_c^{(\mathrm{model})}$ increases weakly and rapidly approaches saturation. In the limit $f\to\infty$, we obtain $T_c^* = 1/4$, and $n_c^* = 1/2$, while the critical compressibility factor tends to the exact analytical value
\begin{equation}
	Z_c^{(\mathrm{model})}(f \to \infty) = 2\ln 2 - 1 \simeq 0.386294.
\end{equation}
The limit $f\gg1$ corresponds to strong local repulsion, which suppresses states with multiple occupancy of the cells; in this limit, the model approaches the behavior of the classical lattice gas~\cite{DKPRS26}. It is interesting to note that the value $Z_c = 0.386$ was also obtained in Ref.~\cite{SL76} for the classical lattice gas in the mean-field approximation, whereas for the classical van der Waals model $Z_c^{vdW} = \frac{3}{8} = 0.375$~\cite{Stanley71}.

Thus, the critical compressibility factor reveals a fundamental limitation of the model under consideration: varying the ratio of local repulsion to global attraction, $f$, does not allow the experimental values of $T_c$, $\rho_c$, and $P_c$ to be matched simultaneously. For the selected set of substances, the agreement improves systematically from H$_2$O to POE9, because the corresponding experimental values of $Z_c^{(\textrm exp)}$ approach the nearly $f$-independent model value $Z_c^{(\mathrm{model})}\simeq0.386$. The trend in the POE5--POE7--POE9 series is particularly interesting; however, establishing its general relation to molecular size or molar mass requires comparison with a broader set of real substances.

\begin{table}[htbp]
	\centering
	\noindent\caption{Comparison of the critical compressibility factor values for the studied substances. Here, $Z_c^{(exp)} = P_c/(n_c k_B T_c)$,
		$Z_c^{(model)} = P_c^*/(n_c^* T_c^*)$}\vskip3mm\tabcolsep4.5pt
	\label{tab_7et}
	
	\noindent{\footnotesize\begin{tabular}{|c|c|c|c|}
			\hline \multicolumn{1}{|c}
			{\rule{0pt}{5mm}Substance} & \multicolumn{1}{|c}{$Z_c^{(exp)}$} &
			\multicolumn{1}{|c}{$Z_c^{(model)}$} &
			\multicolumn{1}{|c|}{$\delta_Z$}\\[2mm]%
			\hline%
			\rule{0pt}{5mm}CO$_2$ & 0.275 & 0.386 & $\sim$29\%\\%
			H$_2$O & 0.229 & 0.386 & $\sim$41\%\\%
			POE5 & 0.309 & 0.386 & $\sim$20\%\\%
			POE7 & 0.320 & 0.386 & $\sim$17\%\\%
			POE9 & 0.347 & 0.386 & $\sim$10\%\\[2mm]%
			\hline
	\end{tabular}}
\end{table}

\begin{table}[htbp]
	\centering
	\noindent\caption{Critical-point coordinates and the critical compressibility factor $Z_c^{(model)}$ for different values of the parameter $f$. The experimental values of the critical compressibility factor $Z_c^{(exp)}$ are 0.275, 0.229, 0.309, 0.320, 0.347 for CO$_2$, H$_2$O, POE5, POE7, POE9, respectively}\vskip3mm\tabcolsep4.5pt
	\label{tab_8et}
	
	\noindent{\footnotesize\begin{tabular}{|c|c|c|c|c|}
			\hline \multicolumn{1}{|c}
			{\rule{0pt}{5mm}$f$} & \multicolumn{1}{|c}{$T_c^*$} &
			\multicolumn{1}{|c}{$n_c^*$} &
			\multicolumn{1}{|c}{$P_c^*$} &
			\multicolumn{1}{|c|}{$Z_c^{(model)}$}\\[2mm]%
			\hline%
			\rule{0pt}{5mm}1.0 & 0.261408 & 0.535499 &
			0.0535261 & 0.382373\\%
			1.00001 & 0.261408 & 0.535498 &
			0.0535258 & 0.382373\\%
			1.0001 & 0.261403 & 0.535482 &
			0.0535234 & 0.382375\\%
			1.001 & 0.261353 & 0.535321 &
			0.0534998 & 0.382393\\%
			1.01 & 0.260870 & 0.533762 &
			0.0532703 & 0.382573\\%
			1.05 & 0.258989 & 0.527745 &
			0.0523838 & 0.383258\\%
			1.1 & 0.257135 & 0.521888 &
			0.0515199 & 0.383916\\%
			1.2 & 0.254567 & 0.513896 &
			0.0503397 & 0.384799\\%
			1.3 & 0.252967 & 0.508981 &
			0.0496136 & 0.385333\\%
			1.5 & 0.251285 & 0.503869 &
			0.0488583 & 0.385883\\%
			1.7 & 0.250567 & 0.501703 &
			0.0485384 & 0.386114\\%
			2.0 & 0.250169 & 0.500507 &
			0.0483616 & 0.386241\\%
			2.5 & 0.250023 & 0.500068 &
			0.0482969 & 0.386287\\%
			3.0 & 0.250003 & 0.500009 &
			0.0482882 & 0.386293\\%
			3.5 & 0.250000 & 0.500001 &
			0.0482870 & 0.386294\\%
			4.0 & 0.250000 & 0.500000 &
			0.0482868 & 0.386294\\%
			5.0 & 0.250000 & 0.500000 &
			0.0482868 & 0.386294\\%
			7.0 & 0.250000 & 0.500000 &
			0.0482868 & 0.386294\\%
			10.0 & 0.250000 & 0.500000 &
			0.0482868 & 0.386294\\[2mm]%
			\hline
	\end{tabular}}
\end{table}

\section{Triple point and the ratio of the critical to triple-point temperatures}
\label{sec8}

Beyond critical characteristics, triple-point coordinates are available for the studied substances. At the triple point, three phases of a substance can coexist in thermodynamic equilibrium at certain temperature and pressure but with different densities. The triple-point parameters for CO$_2$~\cite{SW96},
H$_2$O~\cite{WP02} and POE5, POE7, POE9 \cite{BFHetAl19}, together with the ratio
$T_c/T_{tr}$, are given in Table~\ref{tab_9et}. Only the density of the liquid phase is reported here, whereas those of the gas and solid phases are not included.

For $\mathrm{CO_2}$~\cite{SW96} and $\mathrm{H_2O}$~\cite{WP02}, the triple-point temperature and pressure are taken directly from experimental data, whereas the liquid-phase density at the triple point was calculated using equations of state.
In the case of POE5, POE7, and POE9, the critical-point values $T_c$, $\rho_c$, and
$P_c$ appearing in Table~\ref{tab_1et} were determined in Ref.~\cite{BFHetAl19} by optimizing the equation of state parameters using experimental data.
The triple-point temperature $T_{tr} = 250.0 \, \text{K}$ and pressure
$P_{tr} = 1.0 \times 10^{-12} \, \text{Pa}$ for these lubricants (Table~\ref{tab_9et}) were not determined experimentally.
Rather, they serve as conventional reference parameters of the equation of state. The liquid-phase density $\rho_{tr}^l$ at the triple point is calculated from equations of state constructed individually for each lubricant.

\begin{table*}[htbp]
	\centering
	\noindent\caption{Reference data for the triple-point coordinates and the corresponding values of the ratio $T_c/T_{tr}$}\vskip3mm\tabcolsep4.5pt
	\label{tab_9et}
	
	\noindent{\footnotesize\begin{tabular}{|l|c|c|c|c|c|}
			\hline \multicolumn{1}{|c}
			{\rule{0pt}{5mm}Quantity} & \multicolumn{1}{|c}{CO$_2$} &
			\multicolumn{1}{|c}{H$_2$O} &
			\multicolumn{1}{|c}{POE5} &
			\multicolumn{1}{|c}{POE7} &
			\multicolumn{1}{|c|}{POE9}\\[2mm]%
			\hline%
			\rule{0pt}{5mm}Temperature $T_{tr} \, (\text{K})$ & 216.592 & 273.16 &
			250.0 & 250.0 & 250.0\\%
			Liquid mass density $\rho_{tr}^l \, (\text{kg/m}^3)$ & 1178.53 & 999.793 &
			1053.9 & 1011.8 & 989.8\\%
			Pressure $P_{tr} \, (\text{Pa})$ & $0.51795 \times 10^6$ & 611.657 &
			$1.0 \times 10^{-12}$ & $1.0 \times 10^{-12}$ & $1.0 \times 10^{-12}$\\%
			$T_c/T_{tr}$ & 1.40 & 2.37 &
			3.56 & 3.76 & 3.88\\[2mm]%
			\hline
	\end{tabular}}
\end{table*}

Of particular interest is the ratio of the critical temperature $T_c$ to the triple-point temperature $T_{tr}$,
\begin{equation}
	\label{RT}
	R_T = \frac{T_c}{T_{tr}} = \frac{T_c^*}{T^*_{tr}},
\end{equation}
which varies significantly across different substances. Using the results of Ref.~\cite{KDPRS26}, we show that the present cell model can reproduce a wide range of values for this ratio.

In Ref.~\cite{KDPRS26}, the cell fluid model with Curie--Weiss interaction was modified by introducing an effective temperature-dependent attractive potential. This modification enabled the derivation of a qualitatively new phase diagram for the generalized model, featuring a triple point alongside the critical point.

The parameters $J_1$ and $v$ of the Curie--Weiss cell model were determined in Section~\ref{sec3} using experimental data for the critical points of several substances. In what follows, these parameters are kept fixed and are not readjusted using the triple-point data.

\begin{figure}[htbp]
	\centering
	\includegraphics[width=0.45\textwidth]{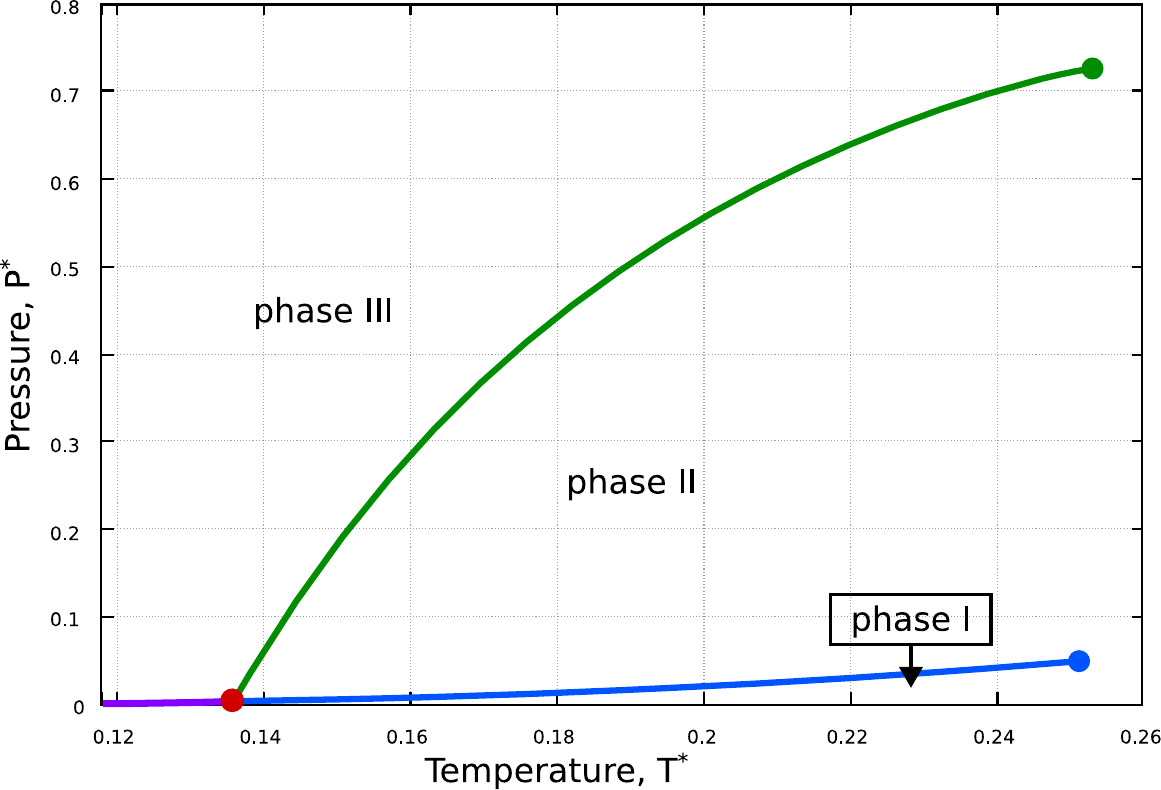}
	\caption{Phase diagram of the cell fluid model with temperature-dependent attraction. The first-order phase-transition lines separating phases~I, II, and III are shown in blue (I--II coexistence), green (II--III coexistence), and violet (I--III coexistence). In the high-temperature region, the blue and green curves terminate at critical points, indicated by circles. The red circle denotes the triple point}
	\label{fig_1et}
\end{figure}

In contrast to the basic model with a temperature-independent attractive interaction, the generalized Curie--Weiss cell model takes into account different temperature dependence of the effective attraction in different phases~\cite{KDPRS26}. It is this feature that leads to the appearance of a triple point in the phase diagram (see Fig.~\ref{fig_1et}, taken from Ref.~\cite{KDPRS26}). In Ref.~\cite{KDPRS26}, the different phases arising in the statistical-mechanical description of the model were denoted as I, II, III, and so on. These phases are characterized by different ranges of their densities. At temperatures below the critical, first-order phase transitions can occur between each pair of successive phases and are characterized by a discontinuous change in density. Phase I has the lowest density, phase II has a somewhat higher density, and so on. In this section, we restrict the consideration to these first three phases. Within a phenomenological comparison with one-component substances, phases I, II, and III can be interpreted as analogues of the gas, liquid, and high-density phases, respectively. The identification of phase III as a solid phase is purely conditional, since the model lacks an order parameter describing crystalline structure.

Each of the three phases labeled by $\pi=I, II, III$, is characterized by its own temperature dependence of the effective attractive interaction:
\begin{eqnarray}
	J_\pi(T^*) & = & J_1 \varphi_{\pi}(T^*), \quad
	\nonumber\\
	\varphi_{\pi}(T^*) & = &
	\begin{cases}
		\left(\frac{T^*}{T^*_c}\right)^{-\alpha_{\pi}},
		& T^* < T^*_c,\\
		1, &  T^* \geq T^*_c.
	\end{cases}
	\label{9d1et}
\end{eqnarray}
At the triple point with temperature $T^* = T_{tr}^*$, all three phases coexist in a state of thermodynamic equilibrium. This means that at $T=T_{\rm tr}$, phases I, II, and III have the same values of the chemical potential and pressure. Accordingly, the three phase-coexistence lines I--II, II--III, and I--III intersect at a single point $(T_{\rm tr},P_{\rm tr})$. The coordinates of the triple point of the generalized model are determined by the choice of the parameters $\alpha_{I}$, $\alpha_{II}$, and $\alpha_{III}$.
For phase I, $\alpha_{I} = 0$ is assumed,
rendering the effective attraction temperature-independent. For phases II and III, positive values of $\alpha_{II}$ and $\alpha_{III}$ are used, satisfying $\alpha_{III} > \alpha_{II}$. As the temperature decreases, the effective attractive interaction in phases II and III becomes stronger, growing faster in phase III than in phase II.

The phase diagram in Fig.~\ref{fig_1et} covers the temperature range from the supercritical region $T^* > T^*_c$ down to the lowest temperature satisfying the stability condition $J_2 > J_\pi(T^*)$. Below this temperature threshold, the attractive interaction becomes dominant over the repulsion, leading to divergence of the integral in the expression for the grand partition function. The critical-point parameters in Fig.~\ref{fig_1et} coincide with those of the model with constant interactions. This is consistent with the assumption for $J_\pi (T^*)$:
upon reaching the critical value $T^*_c$, the function $J_\pi (T^*)$
becomes constant for all $T^* \geq T^*_c$ \cite{KDPRS26}.

By varying the parameters $\alpha_{I}$, $\alpha_{II}$, and $\alpha_{III}$, different values of the ratio $R_T = T_c/T_{tr}$ can be obtained, including values characteristic of real substances. To illustrate this possibility, several values of the parameter $\alpha_{III}$ were considered at fixed $\alpha_{I}$ and $\alpha_{II}$. For each set ($\alpha_{I}$, $\alpha_{II}$, $\alpha_{III}$), the triple-point coordinates $T_{tr}^*$,
$n_{tr}^{*I}$, $n_{tr}^{*II}$, $n_{tr}^{*III}$, and $P_{tr}^*$ were determined, and the ratio $R_T$~\eqref{RT} was also calculated. The obtained results are presented in Table~\ref{tab_10et}.

\begin{table*}[htbp]
	\centering
	\noindent\caption{Triple-point coordinates and the ratio $R_T = T_c/T_{tr}$ for different sets of parameters ($\alpha_{I}$; $\alpha_{II}$; $\alpha_{III}$) at $f = 1.5$ and $T^*_c = 0.2513$}\vskip3mm\tabcolsep4.5pt
	\label{tab_10et}
	
	\noindent{\footnotesize\begin{tabular}{|l|c|c|c|c|c|c|}
			\hline \multicolumn{1}{|c}
			{\rule{0pt}{5mm}($\alpha_{I}$; $\alpha_{II}$; $\alpha_{III}$)} &
			\multicolumn{1}{|c}{$T_{tr}^*$} &
			\multicolumn{1}{|c}{$n_{tr}^{*I}$} &
			\multicolumn{1}{|c}{$n_{tr}^{*II}$} &
			\multicolumn{1}{|c}{$n_{tr}^{*III}$} &
			\multicolumn{1}{|c}{$P_{tr}^* \, (\times 10^{-4})$} &
			\multicolumn{1}{|c|}{$T_c/T_{tr}$}\\[2mm]%
			\hline%
			\rule{0pt}{5mm}(0; 0.1; 0.20) & 0.055950 & 0.000031 &
			0.999969 & 1.999969 & 0.01729048 & 4.49\\%
			(0; 0.1; 0.21) & 0.060762 & 0.000076 & 0.999924 & 1.999924 & 0.04624063 & 4.14\\
			(0; 0.1; 0.22) & 0.065385 & 0.000159 &
			0.999841 & 1.999841 & 0.1038403 & 3.84\\%
			(0; 0.1; 0.23) & 0.069824 & 0.000293 & 0.999708 & 1.999708 & 0.2041818 & 3.60\\
			(0; 0.1; 0.24) & 0.074081 & 0.000491 &
			0.999511 & 1.999511 & 0.3622711 & 3.39\\%
			(0; 0.1; 0.26) & 0.082081 & 0.001114 &
			0.998890 & 1.998890 & 0.9083030 & 3.06\\%
			(0; 0.1; 0.28) & 0.089442 & 0.002076 &
			0.997932 & 1.997925 & 1.836964 & 2.81\\%
			(0; 0.1; 0.30) & 0.096223 & 0.003386 &
			0.996627 & 1.996606 & 3.206628 & 2.61\\%
			(0; 0.1; 0.32) & 0.102482 & 0.005028 & 0.994993 & 1.994944 & 5.039182 & 2.45\\%
			(0; 0.1; 0.34) & 0.108270 & 0.006968 & 0.993062 & 1.992965 & 7.327698 & 2.32 \\[2mm]%
			\hline
	\end{tabular}}
\end{table*}

The data in Table~\ref{tab_10et} show that the triple-point coordinates and the ratio $T_c/T_{tr}$ are sensitive to the choice of parameters that determine the temperature dependence of the attractive interaction. For the considered parameter sets, varying
$\alpha_{III}$ at fixed $\alpha_{I}$ and $\alpha_{II}$ leads to a substantial change in the ratio $T_c/T_{tr}$. Thus, the modification of the model makes it possible, by means of the additional parameter $\alpha_{III}$, to reproduce values of $R_T$ over a broad range of possible experimental values. In particular, for $\alpha_I = 0$ and $\alpha_{II} = 0.1$, the values $0.21 < \alpha_{III} < 0.24$ cover well the range $3.56 \leq R_T \leq 3.88$ for the lubricants POE5, POE7, and POE9 (Table~\ref{tab_9et}). Moreover, values of $\alpha_{III}$ in the interval $0.32 < \alpha_{III} < 0.34$ reproduce the results for water.

It should be noted that the above analysis of the triple point is illustrative in nature. Unlike the critical point, the position of the triple point is not an independent prediction of the model once $J_1$ and $v$ are fixed, as it depends on additional parameters $\alpha_{\rm I}$, $\alpha_{\rm II}$, and $\alpha_{\rm III}$, governing the temperature dependence of the effective attraction. The purpose of our analysis is to show that the generalized model allows the ratio $T_c/T_{\rm tr}$ to be varied over a wide range that includes values characteristic of real one-component fluids. For H$_2$O and CO$2$, the comparison is made with the physical triple points, whereas for POE5, POE7, and POE9 the quoted values of $T_{\rm tr}$ are reference parameters of the corresponding equations of state~\cite{BFHetAl19}.

\section{Conclusions}
In this work, we have evaluated the quantitative applicability of the cell fluid model with Curie--Weiss interaction in describing the critical parameters of real fluids. The analysis was carried out for H$_2$O, CO$_2$, and three polyol ester lubricants -- POE5 (C$_{25}$H$_{44}$O$_8$), POE7 (C$_{33}$H$_{60}$O$_8$) and POE9 (C$_{41}$H$_{76}$O$_8$).

It is shown that the microscopic parameters of the model can be determined from the experimental critical-point coordinates. Using the critical temperature $T_c$ and critical density $\rho_c$, the global attraction constant $J_1$ and the cell volume $v$ are determined uniquely, and the critical pressure becomes an independent prediction of the model. For the studied substances, the calculated critical pressure deviates from the experimental values by $10$?$41\%$.

It is found that varying the ratio of the local repulsion and global attraction constants, $f=J_2/J_1$, over the wide range $1\leq f\leq10$ has only a minor effect on the calculated critical pressure. Consequently, the discrepancy between predicted and experimental value of $P_c$ cannot be eliminated simply by tuning the parameter $f$.

The overall consistency among the three critical-point coordinates ($T_c$, $\rho_c$, and $P_c$) is naturally characterized by the critical compressibility factor $Z_c$. For $f\geq1$, the model gives $Z_c^{(\mathrm{model})}\simeq0.382$?$0.386$, while in the limit $f\rightarrow\infty$ the exact value $Z_c^{(\mathrm{model})}=2\ln2-1\simeq0.386$ is obtained. These values are close to the van der Waals result $Z_c=3/8$ and to the mean-field result for the classical lattice gas, $Z_c=0.386$~\cite{SL76}, and systematically exceed the experimental values. Since matching $T_c$ and $\rho_c$ leads to the relation $P_c^{(\mathrm{model})}/P_c=Z_c^{(\mathrm{model})}/Z_c^{(\mathrm{exp})}$, this discrepancy in $Z_c$ directly determines the overestimation of the model's critical pressure. Varying the parameter $f$ does not resolve this discrepancy.

The best quantitative agreement of the critical pressure and compressibility factor is observed for the polyol ester lubricants, and it consistently improves along the POE5--POE7--POE9 series, whereas the largest deviations are obtained for water. For the set of substances considered, this indicates a tendency toward better agreement of the model with systems composed of large molecules. This tendency is interesting in view of the multiple occupancy of cells and the possible application of the model to soft-matter systems and systems with bounded interaction potentials. To establish a general relationship between the model's predictive accuracy and molecular size, comparison with a broader comparative study covering additional substances is required.

Finally, using the microscopic parameters $J_1$ and $v$ determined at the critical point and the generalization of the model to the case of a phase-dependent temperature-dependent effective attraction, we examine the system's triple-point behavior. It is shown that varying the parameters governing the temperature dependence of the attraction makes it possible to change the position of the triple point and to reproduce a wide range of values of the ratio $T_c/T_{\mathrm{tr}}$ characteristic of the real fluids. No redefinition of the parameters $J_1$ and $v$ fixed at the critical point was required to describe the triple point.

Our findings show both the possibilities and the limitations of the Curie--Weiss cell model for describing real fluids. The model provides a simple relation between the microscopic parameters and the experimental critical-point coordinates and, after introducing a temperature-dependent effective attraction, can be extended to cover the triple-point region. Systematic overestimation of the critical compressibility factor is consistent with the mean-field character of the model and indicates the need for its further refinement.

\vskip3mm \textit{This work was supported by the National Research Foundation of Ukraine under the project No.~2023.03/0201.\\
	The authors are deeply grateful to all warriors of the Ukrainian Armed Forces, living and fallen, for making this research possible.}



\end{document}